\documentclass[traditabstract]{aa}
\usepackage{graphicx}
\usepackage{enumerate}
\usepackage{epsfig}
\usepackage{color}
\usepackage{txfonts}
\usepackage{natbib}

\bibpunct{(}{)}{;}{a}{}{,} 

\begin{document}

\title{Searching for a kinematic signature of the moderately metal-poor stars in the Milky Way bulge using N-body simulations}

\titlerunning{Searching for a kinematic signature of the moderately metal-poor stars in the Milky Way bulge}

\author{A.~G$\rm \acute o$mez\inst{1}\and  P.~Di Matteo\inst{1}\and M.~Schultheis\inst{2}\and F.~Fragkoudi\inst{1}\and M.~Haywood\inst{1}\and F.~Combes\inst{3,4}}

\authorrunning{ G\'omez et al.}

\institute{GEPI, Observatoire de Paris, Universit$\rm\acute e$ PSL, CNRS, 
5 place Jules Janssen, 92190 Meudon, France\\
\email{ana.gomez@obspm.fr}
\and Laboratoire Lagrange, Universit$\rm\acute e$ de la C\^ote d'Azur, Observatoire de la C\^ote d'Azur, CNRS,
Bd de l'Observatoire, 06304 Nice, France
\and
LERMA, Observatoire de Paris, Universit$\rm\acute e$ PSL, CNRS, UPMC, Sorbonne Univ.,
61 Av. de l$'$Observatoire, 75014 Paris, France
\and College de France, 11 Place Marcelin Berthelot, 75005, Paris, France
}

\date{Accepted, Received}

\abstract{Although there is consensus that metal-rich stars in the Milky Way bulge are formed via secular evolution of the thin disc, the origin of their metal-poor counterparts is still under debate. Two different origins have been invoked for metal-poor stars: they might be classical bulge stars or stars formed via internal evolution of a massive thick disc. We use N-body simulations to calculate the kinematic signature given by the difference in the mean Galactocentric radial velocity ($\Delta\rm V_{\rm GC}$) between metal-rich stars ([Fe/H] $\ge$ 0) and moderately metal-poor stars (-1.0 $\le$ [Fe/H] $<$ 0) in two models, one containing a thin disc and a small classical bulge (B/D=0.1), and the other containing a thin disc and a massive centrally concentrated thick disc. We reasonably assume that thin-disk stars in each model may be considered as a proxy of metal-rich stars. Similarly, bulge stars and thick-disc stars may be considered as a proxy of metal-poor stars. We calculate $\Delta\rm V_{\rm GC}$ at different latitudes ($b=0^\circ$, $-2^\circ$, $-4^\circ$,$-6^\circ$, $-8^\circ$ and $-10^\circ$) and longitudes ($l=0^\circ$, $\pm5^\circ$, $\pm10^\circ$ and $\pm15^\circ$) and show that the $\Delta\rm V_{\rm GC}$ trends predicted by the two models are different. We compare the predicted results with  ARGOS data and APOGEE DR13 data and show that moderately metal-poor stars are well reproduced with the co-spatial stellar discs model, which has a massive thick disc. Our results give more evidence against the scenario that most of the metal-poor stars are classical bulge stars. If classical bulge stars exists, most of them probably have metallicities [Fe/H] $<$ -1 dex, and their contribution to the mass of the bulge should be a small percentage of the total bulge mass.}

\maketitle

\section{Introduction}
In recent years, major progress has been made in our understanding of the Galactic bulge. 
Large photometric and/or spectroscopic data sets have allowed us to learn about the physical properties of 
the Galactic bulge components such as the three-dimensional structure, metallicity distribution, kinematics, and
age, but their number and the corresponding mass budget is still under debate. 

From spectroscopic studies of the bulge metallicity distribution function at different galactic longitudes and latitudes, \cite{babusiaux10}, \cite{hill11}, \cite{uttenthaler12}, \cite{rojas14}, \cite{gonzalez15}, \cite{rojas17}, \cite{schul17}, and \cite{zoccali17} found two main stellar components: a metal-rich component centred on [Fe/H] $\sim$ 0.3\,-\,0.4\,dex and a metal-poor component centred on [Fe/H] $\sim$ -0.3\,-\,-0.4\,dex. They display distinct three-dimensional structure and kinematic behaviour.  Metal-rich stars ([Fe/H]$\ge$0 dex) follow the X-shaped structure of a boxy peanut \cite[]{mcwillzoc10,nataf10}, show cylindrical rotation \cite[]{how08, how09, kunder12, ness13b, zoccali14}, and streaming motions \cite[]{rangwala09, depropris11, uttenthaler12, poleski13, babusiaux14, rojas14}, which argue for a  secular formation scenario of the population. Although there is consensus that  the metal-rich stars support the B/P structure and are formed via internal evolution of the thin disc \cite[e.g.][]{combes81,raha91, athanassoula05, debattista06, martinez06, dimatteo14}, the origin of their metal-poor counterparts is controversial. 
Metal-poor stars ([Fe/H]$<$0 dex) do not display the X-shaped structure. They are kinematically different from metal-rich stars. They generally rotate more slowly and are dynamically hotter. They were associated with a classical bulge population, formed via mergers or by a rapid dissipative collapse at early phases of Galaxy formation. However, the ARGOS spectroscopic survey \citep{freeman13} data shed light on another possible origin of low-metallicity stars. Using the full ARGOS sample, \cite{ness13a} identified five different components, A, B, C, D, and E, whose metallicities peak at [Fe/H]~0.1 dex, -0.3 dex, -0.7 dex, -1.2 dex, and -1.7 dex, respectively, the fraction of stars in each component changing with Galactic latitude. 
The most significant components A ([Fe/H]$\ge$ 0), B (-0.5 $\le$ [Fe/H] $<$ 0), and C (-1.0 $\le$ [Fe/H] $<$ -0.5) are consistent with a common disc origin where component C was formed out of the thick disc \cite[]{ness13b, dimatteo15, por17}. \cite{dimatteo14, dimatteo15}, and \cite{dimatteo16} suggest that the B component is associated with the younger thick disc \citep{haywood13}. The more metal-poor components D and E represent less than about 6$\%$ of stars in the ARGOS survey and may be associated with the metal-weak thick disc or with either a possibly classical bulge or inner halo, respectively \cite[]{por17, ger17}. \\
Recently, \cite{bensby17} analysed high-resolution spectra of microlensed dwarf and subgiant stars. A very wide metallicity distribution is observed, ranging from [Fe/H]  $\sim$ -2 to 0.5 dex, with more than two significant peaks. Furthermore, the age distribution is also very wide. For [Fe/H]$>$0 dex, the stars span all ages, from 1 Gyr to 12-13 Gyr. Below [Fe/H]$\le$-0.5 dex, most stars are 10 Gyr or older. 
Moreover, alpha-element abundance trends with metallicity of bulge stars with sub-solar metallicities, even though there are differences, show similarities with the local thick disc. The
authors concluded that the observed age and abundance properties suggest a secular origin for the Galactic bulge, but they cannot
rule out a small contribution of classical bulge or halo stars. On the other hand, studies based on red giants in the bulge also found abundance trends with metallicity to be similar to the nearby thick disc for sub-solar metallicities \citep[e.g.][]{melen08, alves10, ryde10, gonzalez11, john14, gonzalez15, ryde16, jon17, garciaperez18}.\\
The metallicity distribution observed from dwarfs and giants in the bulge displays less than about 5-6$\%$ of metal-poor stars with [Fe/H]$<$-1.0 dex; most of them have [Fe/H] values $\ge$-1.0 dex. In what follows, stars with -1.0 $\le$ [Fe/H] $<$ 0 are called moderately metal-poor stars, and those with [Fe/H] $\ge$ 0 are called metal-rich stars. Stars with [Fe/H]$<$-1.0 dex are well traced by RRLyrae stars, whose origin is also controversial. \cite{kun16} suggested that the observed spatial distribution and kinematics are consistent with a classical bulge origin, although they cannot rule out the possibility that they are the metal-poor tail of a more metal-rich halo-bulge population. Instead, \cite{per17} showed that RRLyrae stars in the bulge might be the inner extension of the Galactic stellar halo.

We here investigate whether the origin of moderately metal-poor stars (-1.0 $\le$ [Fe/H] $<$ 0) is mainly consistent with a classical bulge origin or with a formation from a rather massive thick disc present in the inner Galaxy. Using N-body simulations, we have calculated the difference in the mean Galactocentric radial velocity ($\Delta\rm V_{\rm GC}$) between metal-rich and metal-poor stars in two models; one containing a small classical bulge, and the other consisting of only disc components. The kinematic signature given by each of the two models was obtained at different lines of sight. The predicted kinematic trends were compared with data from the ARGOS survey \citep{freeman13} and from the APOGEE DR13 survey \cite[]{SDD17, maj17}. The results show that the observed data are consistent with a thick-disc origin for moderately metal-poor stars, leaving little room for a classical bulge origin.

The paper is organised as follows: in Section 2, the models are briefly described. Section 3 shows the kinematic results of the simulations at different latitudes and longitudes.  The comparison to observations is shown in Section 4. In Sections 5 and 6, we discuss and summarise our results, respectively.

\section{Simulations}
Two different high-resolution simulations are considered in this paper. The first, called the ``disc+bulge" here, consists of an isolated stellar disc and a classical bulge, with a bulge-to-disc ratio equal to B/D=0.1, and contains no gas. This simulation has been extensively described and analysed in \cite{dimatteo14}, \cite{dimatteo15}, and \cite{gom16}. 
The dark halo and the bulge are modelled as Plummer spheres \citep{BT87}. The dark halo has a mass $M_H = 1.02 \times 10^{11} M_{\odot}$ and a characteristic radius $r_H=10$~kpc. The bulge has a mass $M_B = 9 \times 10^9M_{\odot}$ and characteristic radius $r_B=1.3$~kpc. The stellar disc follows a Miyamoto-Nagai density profile \citep{BT87}, with mass $M_* = 9 \times 10^{10} M_{\odot}$ and vertical and radial scale lengths given by $h_* =0.5$~kpc  and $a_* =4$~kpc, respectively. The initial disc size is 13~kpc, and the Toomre parameter is set equal to Q=1.8. The galaxy is represented by 30\,720\,000 particles redistributed among dark matter (10\,240\,000) and stars (20\,480\,000). This simulation reproduces the observed trends of the global kinematics and the global chemical characteristics of the Galactic bulge. However, it does not reproduce the chemo-kinematic relations satisfied by the individual bulge components. \cite{dimatteo15}, assuming an initial radial metallicity profile in the disc similar to \cite{martinez13}, showed that this model fails in reproducing the observed properties of stellar components in ARGOS data (see their Fig. 4).

That the Galactic thick disc seems to be as massive in the inner Galaxy as the Galactic thin disc \citep{fuhrmann12, haywood13, snaith14} led \cite{dimatteo14} to suggest that the Milky Way bulge is the result of mapping the Galactic thin+thick disc into the boxy/peanut-shaped structure. Thus, we have considered a second simulation, called ``thin+thick" here, which consists of three isolated stellar discs corresponding to a kinematically cold thin disc, to an intermediate disc with intermediate kinematics, and to a kinematically hot disc (hereafter called ``thin", ``intermediate", and ``thick" discs). They have different scale heights and lengths, with masses and sizes in agreement with recent estimates for the Milky Way \cite[]{bensby11, bovy12a, bovy12b, bensby13, haywood13, snaith14, hayden15, bovy16}. The discs are modelled with Miyamoto-Nagai density distributions \citep{BT87} with masses $M_* = 2.55 \times 10^{10} M_{\odot}$, 1.53$\times 10^{10} M_{\odot}$, and $1.02\times 10^{10} M_{\odot}$; scale heights $h_* =0.3$~kpc, 0.6~kpc, and 0.9~kpc; and scale lengths $a_* =4.7$~kpc, 2.3~kpc, and 2.3~kpc for the thin-disc, the intermediate-disc, and the thick-disc components, respectively. The dark halo is modelled as a Plummer sphere \citep{BT87} with characteristic mass and radius $M_H = 1.61\times 10^{11} M_{\odot}$ and  $r_H=10$~kpc. The modelled disc galaxy consists of 25\,000\,000 particles, redistributed among stars (20\,000\,000) and dark matter (5\,000\,000). Of the star particles, 10\,000\,000 belong to the thin disc, 6\,000\,000 to the intermediate disc, and 4\,000\,000 to the thick disc. The stellar disc is structured with scale heights and velocity dispersions at the solar vicinity similar to those observed for the thin disc, the young thick disc, and the old thick disc, respectively \citep{haywood13}. The intermediate and thick discs together represent 50\% of the stellar mass, the remaining 50\% being in the thin disc. All stellar components contribute to the stellar bar and to the boxy/peanut-shaped structure, but their fraction varies with the height above the plane. Thin-disc stars are more concentrated towards the galactic plane, the fraction of thick-disc stars increases with height above the plane, and that of intermediate-disc stars stays nearly constant \citep{dimatteo16}, with proportions similar to those for populations A, B, and C by \cite{ness13a}. At a given height above the plane, the boxy/peanut-shaped structure is more pronounced in the kinematically cold populations than in the hottest population \citep{dimatteo16, frag17}. Results on the kinematic characteristics predicted by this disc model are given in Di Matteo et al. (in preparation).

In an N-body simulation the length of the bar is difficult to control, and it is therefore necessary to rescale the model to match the length of the bar of the Milky Way.
Both models were rescaled to match the Milky Way bar size and bulge velocities; see \cite{dimatteo15} for the disc+bulge simulation and Di Matteo et al. (in preparation) for the thin+thick simulation, the Sun being placed  at 8 kpc from the Galactic centre.
For consistency, we adopted a bar orientation relative to the Sun-Galactic centre of 20$^\circ$ as in \cite{dimatteo15} and \cite{dimatteo16}. As a consequence of the observed three-dimensional structure of the Milky Way box/peanut-bulge \citep[e.g.][]{dwek95, cao13, wegg13, bland16}, its extent in the y-direction is larger than in the x-direction, where x and y are measured on the Galactic plane from the Galactic centre, y in the direction Sun-Galactic centre, and x perpendicular to it. To avoid contamination from background and foreground stars, we selected stars inside $|x|\le 2.5$~kpc and $|y|\le 3$~kpc. The extent in the z-direction, the axis perpendicular to the Galactic plane, depends on the selected galactic latitude line of sight; it is of about 2~kpc at $|b|=10^\circ$. \\

The spatial structure of stellar populations with different chemical abundances in the Milky Way stellar disc is complex. \cite{bovy12a} showed that the Milky Way disc has a continuum of stellar mono-abundance populations, each having a simple spatial structure in both the vertical and the radial directions. \cite{hayden15} also showed that the stellar abundance distribution in the Galactic disc varies with cylindrical Galactocentric distance ($R$) and with height $|z|$ (see their Fig. 4). In particular, stars with high-[Fe/H] values ([Fe/H] $>$ 0 dex) are more confined to the mid-plane and to the inner disc (3~$< R <$ 5 ~kpc). At least in the inner disc, the metal-poor mono-abundance populations have greater scale-heights, while the metal-rich mono-abundance populations have smaller scale-heights \citep{bovy16}. Because inner thin-disc stars are metal-rich and because we studied the mapping of the inner disc into the B/P bulge,
we reasonably assumed that the disc stars in the simulation disc+bulge and the thin-disc stars in the simulation thin+thick may be considered as a proxy of metal-rich stars. On the other hand, bulge stars and intermediate-disc and thick-disc stars may be considered as a proxy of metal-poor stars.
We calculated the difference in the mean Galactocentric radial velocity between metal-rich and metal-poor stars ($\Delta\rm V_{\rm GC}$ = $V_{\rm GC}$(metal-rich) - $V_{\rm GC}$(metal-poor)) in both models at different latitudes ($b=0^\circ$, $-2^\circ$, $-4^\circ$,$-6^\circ$, $-8^\circ$ and $-10^\circ$), and longitudes ($l=0^\circ$, $\pm5^\circ$, $\pm10^\circ$ and $\pm15^\circ$). The $\Delta\rm V_{\rm GC}$ trends predicted by the two models are different (see next section), and when compared to observations, they may help to shed some light on the origin of moderately metal-poor stars.

\begin{figure*}
\centering
\includegraphics[width=6cm,angle=0]{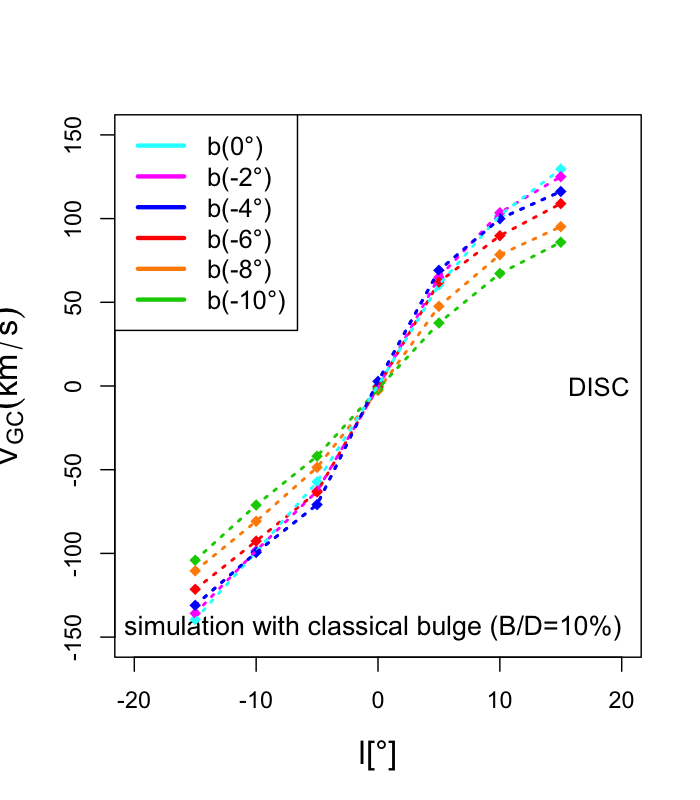}
\includegraphics[width=6cm,angle=0]{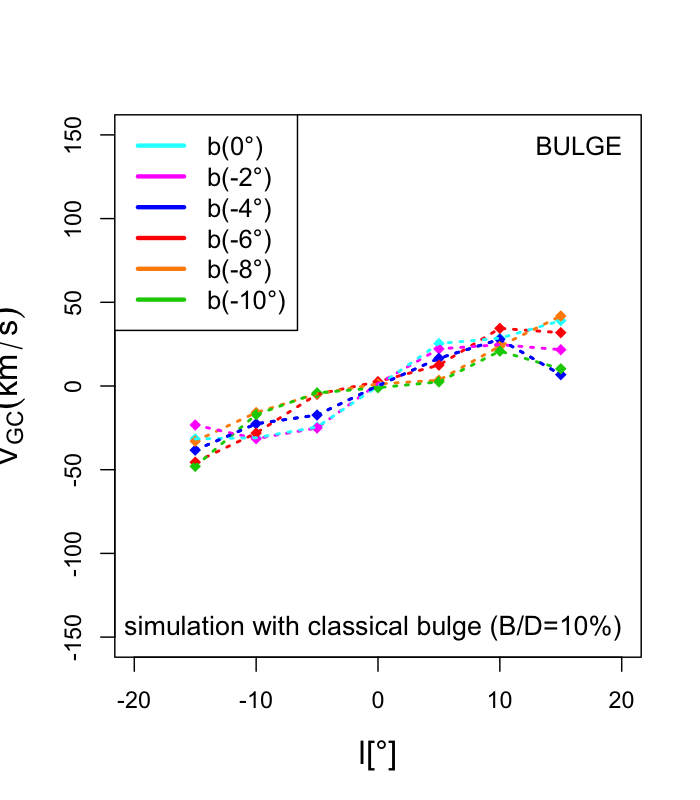}
\includegraphics[width=6cm,angle=0]{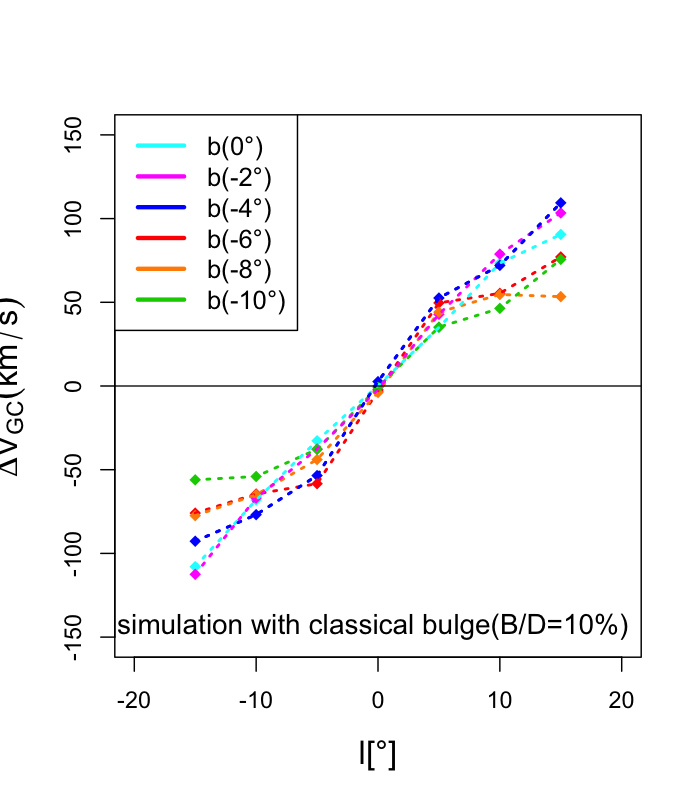}
\caption{Rotation curve of disc stars (\textit{left panel}) and of classical bulge stars (\textit{middle panel)} of the $N$-body model with a disc and a classical spheroid (B/D=10\%) (disc+bulge simulation). $\Delta\rm V_{\rm GC}$ = $V_{\rm GC}$(disc) - $V_{\rm GC}$(classical bulge) between disc stars and classical bulge stars (\textit{right panel}). Stars at $|x|\le 2.5$~kpc and $|y|\le 3$~kpc from the Galactic centre are selected. Different latitudes are shown for the modelled galaxy: $b=0^\circ$ , $-2^\circ$, $-4^\circ$,$-6^\circ$, $-8^\circ$, and $-10^\circ$. The size of the fields is $ \Delta{l}=\Delta{b} =1^\circ$ for $b$ ${\geq}-6^\circ$. To obtain better number statistics for samples at latitudes $b$ $<-6^\circ$, $ \Delta{l}=\Delta{b} =1^\circ.5$ have been adopted.}
\label{VGC_DEL_disc_bulge}
\end{figure*}

\section{Results}
Fig.~\ref{VGC_DEL_disc_bulge}  shows the rotation curve as deduced by radial velocity measurements in the model with a spheroid for the disc and the bulge components separately, as well as the difference in the mean Galactocentric radial velocity $\Delta\rm V_{\rm GC}$ between disc and classical bulge stars. In Fig.~\ref{VGC_DEL_disc_bulge}, the size of the fields is $ \Delta{l}=\Delta{b} =1^\circ$ for $b$ ${\geq}-6^\circ$ and $ \Delta{l}=\Delta{b} =1^\circ.5$  at latitudes $b$ $<-6^\circ$.

As shown in \cite{dimatteo14}, the disc component displays approximately cylindrical rotation with a weak  dependence of rotation speed on latitude; the velocity rotation curve is steeper towards lower latitudes. This effect has previously been noted by \cite{how09} in BRAVA data. The difference in the velocity curves with latitude is similar to what is observed in ARGOS data \citep{ness13a} and in GIBS data \citep{zoccali14}.  Even though classical bulge stars acquire some angular momentum during the bar formation and evolution \cite[]{saha12, saha13}, their rotational velocities remain much lower than those of disc stars. This is clearly shown in Fig.~\ref{VGC_DEL_disc_bulge} (right panel), where $|\Delta\rm V_{\rm GC}|$ values are higher than about 50 $\rm km$ $\rm s^{-1}$ for $|l|\ge \sim 5^\circ$.

\begin{figure*}
\centering
\includegraphics[width=6cm,angle=0]{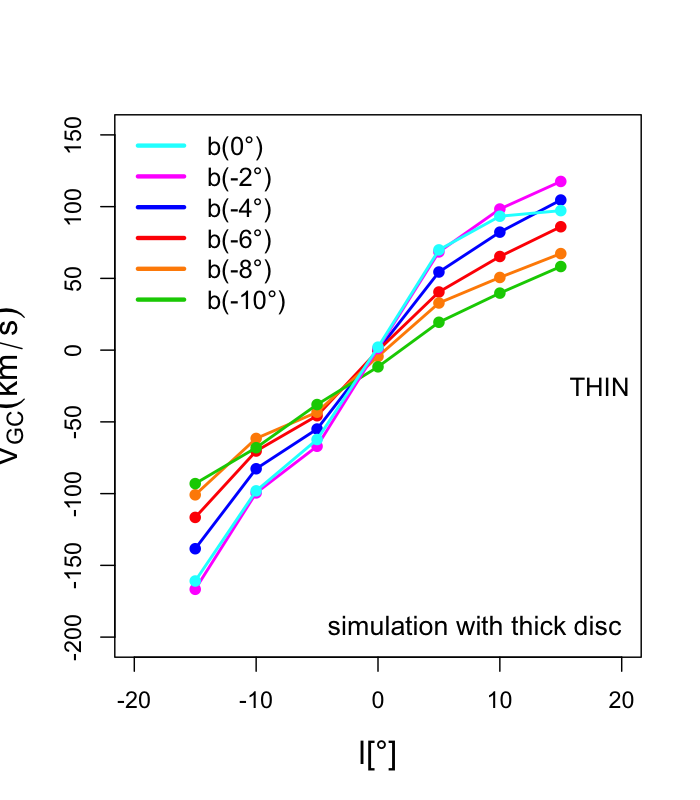}
\includegraphics[width=6cm,angle=0]{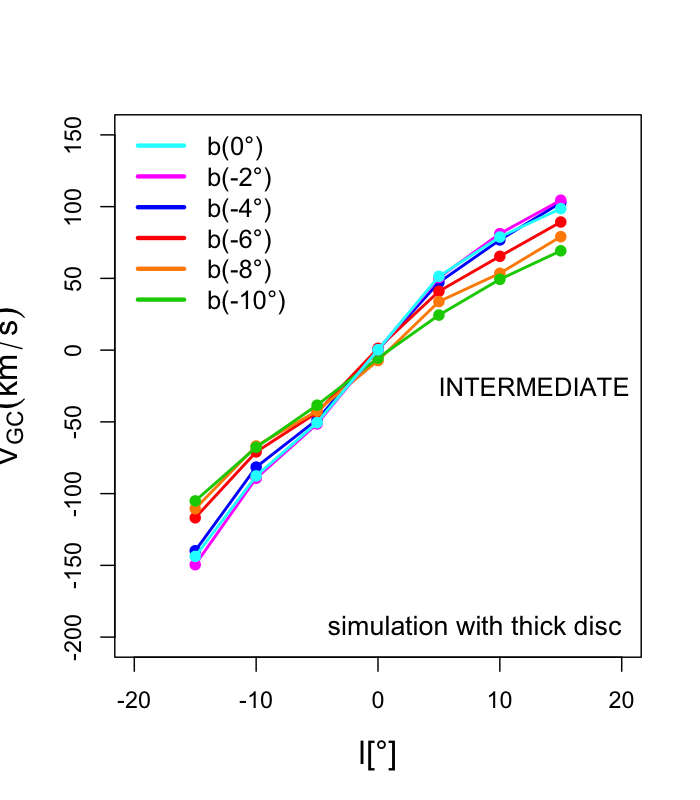}
\includegraphics[width=6cm,angle=0]{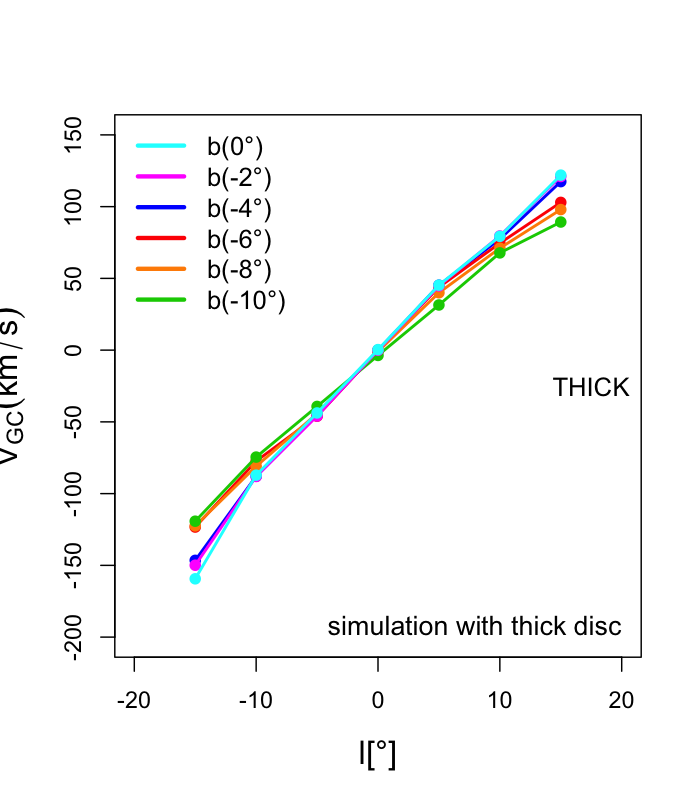}
\caption{Rotation curve of thin-disc stars (\textit{left panel}), intermediate-disc stars (\textit{middle panel}) and thick-disc stars (\textit{right panel}) of the $N$-body disc galaxy model (thin+thick simulation). Stars at $|x|\le 2.5$~kpc and $|y|\le 3$~kpc from the Galactic centre are selected. Different latitudes are shown for the modelled galaxy: $b=0^\circ$ , $-2^\circ$, $-4^\circ$,$-6^\circ$, $-8^\circ$, and $-10^\circ$. The size of the fields is $ \Delta{l}=\Delta{b} =1^\circ.5$. }
\label{VGC_DE}
\end{figure*}

Fig.~\ref{VGC_DE} shows the rotational velocity for the three disc components in the thin+thick model. The modelled disc is made of 20\,000\,000 particles, redistributed in the three components, compared to about 18\,800\,000 particles of the disc in the disc+bulge model. In order to obtain better number statistics, the adopted size of the fields was $\Delta{l}=\Delta{b} =1^\circ.5$. We recall that thin-disc stars are concentrated towards the Galactic plane, and they constitute 40$\%$ of the total stellar density at $|b| < \sim 2^\circ$ and contribute to less than 25$\%$ at $|b| \sim 8^\circ -10^\circ$ \citep{dimatteo16}. The opposite trend is observed for thick-disc stars: their fraction increases with galactic latitude from about 20$\%$ at $|b| < \sim 2^\circ$ to more than 40$\%$ at $|b| \sim 8^\circ -10^\circ$ \citep{dimatteo16}. Fig.~\ref{VGC_DE} shows that the hottest component (thick disc) exhibits cylindrical rotation with only weak latitude variations. The thin-disc component does not show cylindrical rotation for all latitudes, the rotation velocities being similar up to $b\ge-2^\circ$ and $|l|\le \sim 5^\circ$. The intermediate-disc component displays cylindrical rotation up to about $b\ge-4^\circ$, but the variation in rotational velocity as a function of the latitude is smaller than that corresponding to thin-disc stars. Our results are similar to those predicted by the M2M chemodynamical model of \cite{por17} based on the ARGOS data and the APOGEE DR12 data: metal-rich stars (bin A, 0.5 $\ge$ [Fe/H] $\ge$ 0.0) and metal-poor stars (bin B, 0.0 $\ge$ [Fe/H] $\ge$-0.5) do not exhibit cylindrical rotation for all latitudes (see their Fig. 15).

\begin{figure*}
\centering
\includegraphics[width=6cm,angle=0]{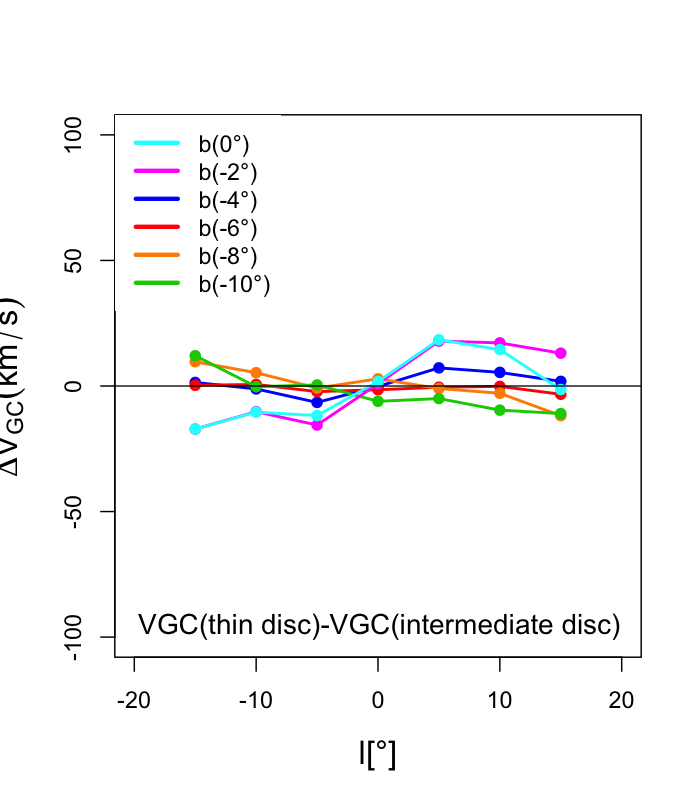}
\includegraphics[width=6cm,angle=0]{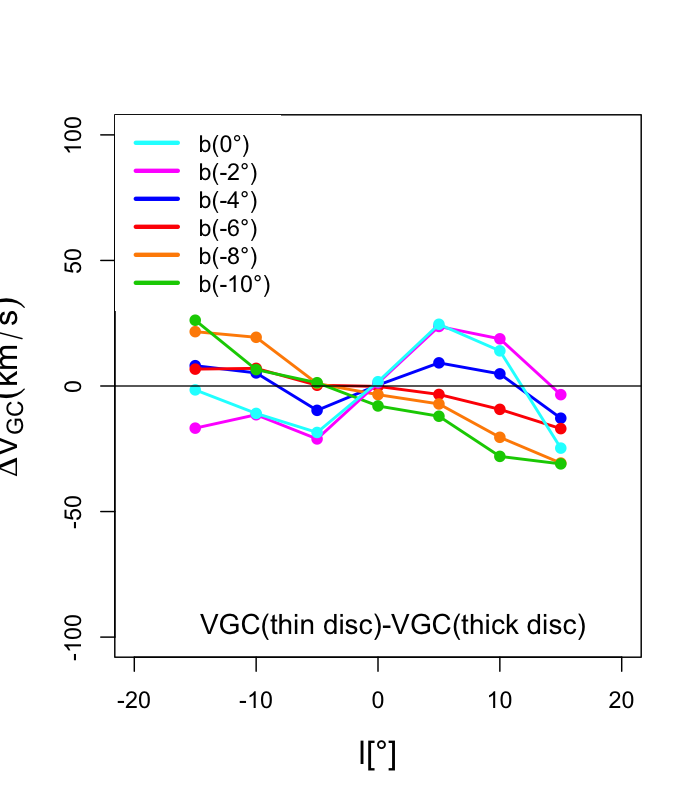}
\includegraphics[width=6cm,angle=0]{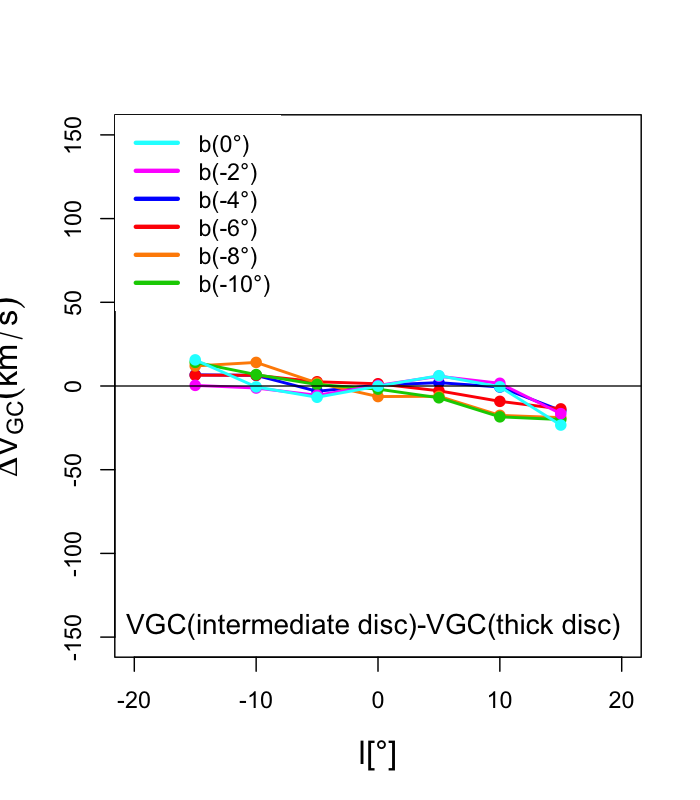}
\caption{$\Delta\rm V_{\rm GC}$ values between stars in the different disc components in the thin+thick simulation. 
$V_{\rm GC}$(thin disc) - $V_{\rm GC}$(intermediate disc) (\textit{left panel}); $V_{\rm GC}$(thin disc) - $V_{\rm GC}$(thick disc) (\textit{middle panel}), and $V_{\rm GC}$(intermediate disc) - $V_{\rm GC}$(thick disc) (\textit{right panel}). Stars at $|x|\le 2.5$~kpc and $|y|\le 3$~kpc from the Galactic centre are selected. Different latitudes are shown: $b=0^\circ$ , $-2^\circ$, $-4^\circ$,$-6^\circ$, $-8^\circ$, and $-10^\circ$. The size of the fields is $\Delta{l}=\Delta{b} =1^\circ.5$.}
\label{VGC_DEL_DE}
\end{figure*}

Fig.~\ref{VGC_DEL_DE} shows the mean radial velocity differences $\Delta\rm V_{\rm GC}$ between the modelled disc components. Intermediate-disc stars rotate slightly faster than thin-disc stars (left panel) for $b < \sim -4^\circ$, while 
for $b > \sim -4^\circ$, the trend is inverse, the maximum difference in $|\Delta\rm V_{\rm GC}|$ being of about 10 $\rm km$ $\rm s^{-1}$. \cite{por17} found a similar trend for stars in bins A and B and argued that it is likely the result of the complex orbital structure of the bar in the plane.
Like intermediate-disc stars, thick-disc stars also display faster rotation than thin-disc stars (middle panel) only for $b < \sim -4^\circ$, but the maximum difference $|\Delta\rm V_{\rm GC}|$ is of about 20 $\rm km$ $\rm s^{-1}$. Recently, \cite{frag17}, using N-body simulations with a thin disc and a thick disc, found that the rotational velocity of thick-disc stars in the outer parts of the bulge can be larger than those of thin-disc stars by about 20$\%$ for Milky Way-like orientation of the bar. This was interpreted as a consequence of the orbital structure of the thin-disc and thick-disc bars. 
The difference $\Delta\rm V_{\rm GC}$ between intermediate-disc and thick-disc stars (right panel) shows that thick-disc stars rotate slightly faster than the intermediate-disc stars for almost all latitudes. 
Fig.~\ref{VGC_DEL_disc_bulge} (right panel) and Fig.~\ref{VGC_DEL_DE} show that the trends of $|\Delta\rm V_{\rm GC}|$ are different. In the disc+bulge model, at $|l|\le \sim 5^\circ$ $|\Delta\rm V_{\rm GC}|$ is larger than about 40 $\rm km$ $\rm s^{-1}$,
while in the thin+thick model, $|\Delta\rm V_{\rm GC}|$ between the thin and the thick components is smaller than about 20 $\rm km$ $\rm s^{-1}$. 
At $|l|> \sim 5^\circ$, $|\Delta\rm V_{\rm GC}|$ increases in both models, but only slightly in the thin+thick model compared to the disc+bulge model. 
As quoted before, we assume in what follows that disc stars in the disc+bulge simulation and thin-disc stars in the thin+thick simulation may be considered as a proxy of metal-rich stars. Similarly, bulge stars and intermediate-disc and thick-disc stars may be considered as a proxy of moderately metal-poor stars. The different trends observed in both simulations are compared in the next section with the kinematics of some of the available observational data of the Milky Way bulge.

\section{Comparison with observations}
We compared the $\Delta\rm V_{\rm GC}$ results of the two models with data from the ARGOS survey \citep{freeman13} and from the APOGEE DR13 survey \citep{SDD17, maj17} along the lines of sight considered in this paper. Both surveys provide kinematic and metallicity measurements.

\begin{figure*}
\centering
\includegraphics[width=8cm,angle=0]{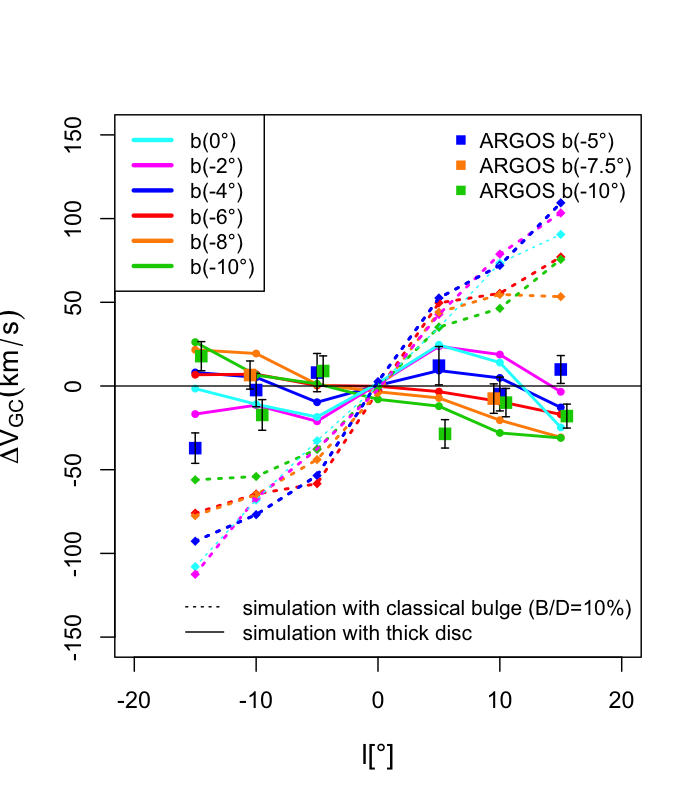}
\includegraphics[width=8cm,angle=0]{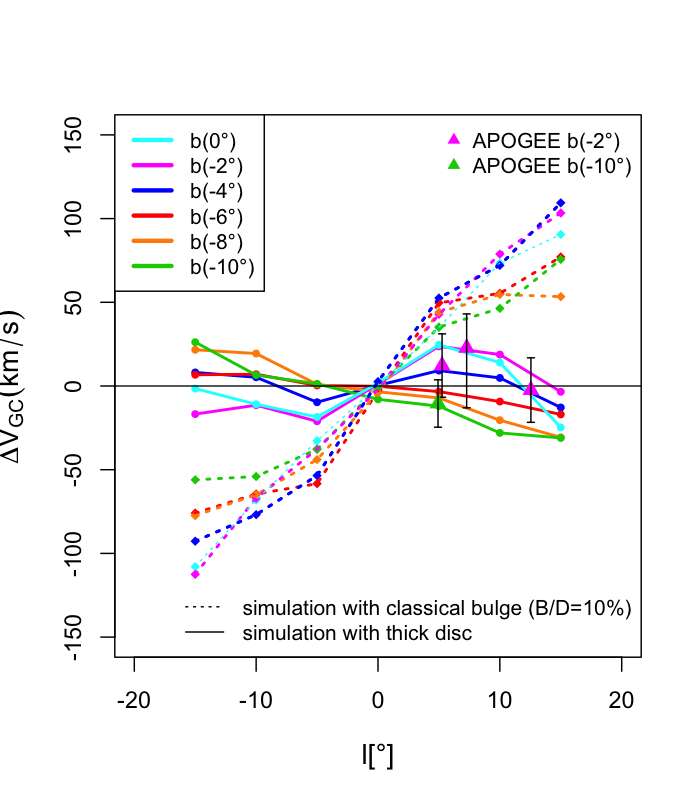}
\caption{Comparison of $\Delta\rm V_{\rm GC}$ values for the disc+bulge and the thin+thick simulations with ARGOS data:  $V_{\rm GC}$(A~component) - $V_{\rm GC}$(C~component) (\textit{left panel}) and APOGGE data $V_{\rm GC}$([M/H]$>$0) - $V_{\rm GC}$([M/H]$<$0) (\textit{right panel}). For details see the text.}
\label{VGC_DEL_ARGOS_APO}
\end{figure*}

The ARGOS survey is a large medium-resolution spectroscopic survey that covers $l$ between ${-15^\circ}$ and ${15^\circ}$ and $b$ between ${-5^\circ}$ and ${-10^\circ}$. Of the metallicity components identified in the survey (A to E) \citep{ness13a}, A has a thin-disc origin, while C is likely to be the old-thick component that
was in place before the bar formation \cite[]{dimatteo14, dimatteo15}. We do not have individual ARGOS data but only average radial velocities for the A, B, and C components at different longitude and galactic latitudes selected within $R \le $ 3.5~kpc \citep{ness13b}. The difference $\Delta\rm V_{\rm GC}$ between the metal-rich component (A) and the metal-poor component (C) is shown in Fig.~\ref{VGC_DEL_ARGOS_APO} (left panel) and is compared with $\Delta\rm V_{\rm GC}$ values of the simulation with classical bulge (dotted lines, Fig.~\ref{VGC_DEL_disc_bulge}, right panel) and the values of the simulation with pure disc components (solid lines, Fig.~\ref{VGC_DEL_DE}, middle panel). ARGOS data are in good agreement with the thin+thick model, supporting the result that the metal-poor component C is associated with the thick disc  \citep[]{ness13a, dimatteo14, dimatteo15, dimatteo16} at the latitudes covered by the ARGOS survey. A similar conclusion has been obtained by \cite{por17} for the C component outside the central kpc.

APOGEE data allow us to complement ARGOS data at latitudes near the Galactic plane.  The APOGEE survey is a high-resolution spectroscopic survey in the near-infrared of Milky Way stellar populations. It provides both accurate radial velocities and reliable abundance measurements. Moreover, it allows us to observe dust-obscured regions of the Galactic bulge. We refer to \cite{maj17} for a complete description of APOGEE, to \cite{zasowski13} for the target selection, and to \cite{garciaperez16} for the ASPCAP: The APOGEE Stellar Parameter and Chemical Abundances Pipeline. To isolate the stars in the bulge region, we selected stars in a range of heliocentric distances between 4 and 12~kpc. Then we removed stars that have log$g> 0.5$ in order to reduce the metallicity bias against metal-rich stars in the inner Galaxy \citep[]{hayden15, ness16}. The distances were estimated using the Padova set of isochrones \citep{schul17}. For each star, the closest point on the isochrones was obtained from the stellar parameters [M/H], log$g$, and T$_{\rm eff}$. Finally, we kept stars with relative error in the distance smaller than 50$\%$, although this cut does not change the results given that most of the stars have a relative error smaller than 40$\%$. The distance selection of APOGEE stars is rather similar to the selection of ARGOS stars (distances between 4.5 and 11.5 ~kpc from the Sun, \citealt{ness16}), taking into account uncertainties in the star distance determinations in both surveys.
Fig.~\ref{VGC_DEL_ARGOS_APO} (right panel) displays the results for four regions: one centred at $l\sim 5^\circ$ and $b\sim -10^\circ$ (110 stars),  and the other three at low Galactic latitude ($b\sim -2^\circ$) and $l\sim 5.2^\circ$ (110 stars), $7.2^\circ$ (91 stars) and $12.5^\circ$ (137 stars), respectively. Each sample was divided into metal-rich stars ([M/H]$\ge$ 0) and metal-poor stars ([M/H]$<0$), and [M/H] represents the global metallicity, which is closely correlated with the iron abundance. There is no systematic difference between [M/H] and [Fe/H] in the APOGEE data \citep{holtz15, garciaperez16}.
$\Delta\rm V_{\rm GC}$ was measured as the difference in the mean radial Galactocentric velocity of metal-rich stars minus metal-poor stars. Like the ARGOS data, the APOGEE data show a trend similar to the observed in the thin+thick model. This result suggests that most of the moderately metal-poor stars in the considered APOGEE samples are compatible with being predominantly thick-disc stars.

\section{Discussion}
The kinematic trends obtained in Section 4 show that moderately metal-poor stars observed in the bulge region (-1.0 $\le$ [Fe/H] $<$ 0) in the ARGOS and in the APOGEE DR13 surveys are well reproduced with the co-spatial stellar discs model, which has a massive thick disc. 
It should be noted that \cite{will16} found from the analysis of the kinematics of Gaia-ESO survey stars \citep{gil12} ``no discernible difference between the rotational signature" in the metal-rich stars ([M/H]$\ge$ 0) and the metal-poor ones ([M/H]$<$ 0). On the other hand, from GIBS low spectral resolution data, \cite{zoccali17} showed that the difference in the Galactic radial velocity between metal-poor and metal-rich components seems marginal.

Elemental abundance trends for metal-poor stars are similar to those of local thick-disc stars. 
Studies based on red giants in the bulge found no evidence for a different behaviour of the alpha-elements versus iron trends in the bulge compared to the local thick disc for sub-solar metallicities \citep[e.g.][]{melen08, alves10, ryde10, gonzalez11, john14, gonzalez15, jon17}. 
\cite{bensby17} observed in their analysis of micro-lensed dwarf and sub-giant stars that the alpha-elements (Mg, Si, Ca, and Ti) trends with [Fe/H] for metal-poor metallicities tend to follow the same trends as can be seen for local thick-disc stars and that the level of alpha-elements enhancement appears slightly higher in the bulge than in the thick disc. They argued that the observed offset would indicate that the star formation rate was slightly faster in the bulge than in the local thick disc, which is consistent with the much denser environment in the inner parts of the Galaxy.  On the other hand, \cite{rojas17} obtained spectroscopic data of red clump stars from the fourth internal data release of the Gaia-ESO survey and found that for metal-poor stars, the [Mg/Fe] vs [Fe/H] trend is comparable to the corresponding to the thick-disc sequence.
 \cite{ryde16} investigated the Galactic centre region at a projected Galactocentric distance of about 300 pc by observing [alpha/Fe] element trends of Mg and Si versus metallicity from high-resolution spectroscopic data for M giants. They found a wide range in metallicities from -1.2 $<$ [Fe/H] $<$ 0.3. Their Fig. 4 shows that the obtained trends are rather similar, within the uncertainties, with the abundance trends based on micro-lensed dwarfs \citep{bensby13} for the outer bulge. 
 
\cite{bensby17} also analysed light odd-Z elements (Na and Al), iron-peak elements (Cr, Ni and Zn), and neutron-capture elements (Y and Ba). With the exception of the [Al/Fe] trend, which appeared to be placed at the upper envelope of the thick-disc trend, the other abundance-[Fe/H] trends display similarities between the thick disc and the metal-poor bulge. From high-resolution spectra, \cite{vanders16} analysed heavy elements (Ba, La, Ce, Nd, and Eu) of bulge stars, most of which have [Fe/H] $\ge$ -1.0 dex. In particular, for metal-poor stars, they found that Ba and La are enhanced with respect to their thick-disc counterparts, but the observed [Eu/Fe] trend is in good agreement with the trend of Galactic thick disc derived by \cite{bensby05} and \cite{reddy06}. Unlike the work of \cite{bensby17}, which is based on homogeneous data sets analysed in a consistent manner,  \cite{vanders16} compared their determinations with that of the literature, which might explain the observed discrepancies.

Recently, a similar thin+thick model has allowed us to reproduce the metallicity gradient in the disc as derived by APOGEE DR13 data \citep{frag17b}. Moreover, the central concentration of metal-poor stars in the inner kpc of the Galaxy \citep[e.g.][]{zoccali17, por17} and the metallicity distribution function of the bulge are also reproduced by the model \citep{frag18}.

To summarise, both kinematic and chemical results  agree with the fact that moderately metal-poor stars are formed out of the thick-disc stars via secular evolution, without the necessity of a massive classical spheroid. However, this does not mean that the Milky Way bulge does not contain classical bulge stars, but that its mass contribution probably is a small percentage of the total bulge mass, as has been reported in several works \citep[e.g.][]{shen10, kunder12, dimatteo14, kun16,debat17}. \cite{howes15} found extremely metal-poor stars ( [Fe/H] $\le$ -2.3 dex) in the bulge. The observed stars follow tight orbits around the Galactic centre and have chemical compositions consistent with typical halo stars of the same metallicity, although they do not have the large carbon enhancements expected in halo stars. Whether most of the stars with [Fe/H] $\le$ -1.0 dex are halo stars or classical bulge stars remains an open question and is
beyond the scope of this paper.

\section{Concluding remarks}
The observed metallicity distribution function of Milky Way bulge stars show that most of them have [Fe/H]$>$-1 dex. We here explored the origin of moderately metal-poor stars (-1.0 $\le$ [Fe/H] $<$ 0) by comparing a kinematic signature given by the difference in the mean Galactocentric radial velocity between metal-rich and metal-poor stars in two different N-body models with  ARGOS data and APOGEE DR13 data. One model consists of a disc and a small classical bulge (B/D = 0.1), and the other consists of a composite stellar disc (a kinematically cold thin-disc, an intermediate disc with intermediate kinematics, and a kinematically hot thick-disc) where the intermediate and thick discs together represent 50\% of the stellar mass.  As our models do not have chemical information, we reasonably assumed that the thin disc in both simulations may be considered as a proxy of metal-rich stars. Similarly, classical bulge stars and intermediate-disc and thick-disc stars may be considered as a proxy of metal-poor stars. The kinematic signature obtained at different latitudes and longitudes shows that the trends predicted by the models are different. We show that the observed kinematic trends from ARGOS data and APOGEE DR13 data are well reproduced with the co-spatial stellar discs model, which argues that the moderately metal-poor stars were formed from the thick-disc stars via secular evolution, leaving little room for a classical bulge origin. Our results give further evidence against the scenario that most of the metal-poor stars are classical bulge stars. If classical bulge stars exist, most of them probably have metallicities [Fe/H]$<$-1 dex, and their contribution to the mass of the bulge should be a small percentage of the total bulge mass.

\section*{Acknowledgments}
This work has been supported by the ANR (Agence Nationale de la Recherche) through the MOD4Gaia project (ANR-15-CE31-
0007, P.I.: P. Di Matteo). FF is supported by a postdoctoral grant from the Centre National d'Etudes Spatiales (CNES).

\end{document}